\documentstyle[prl,multicol,epsf,aps]{revtex}
\draft

\title{Fractional Langevin equation} 
\author{Eric Lutz} 
\address{D\'epartement de Physique Th\'eorique, Universit\'e de Gen\`eve, 24, quai Ernest Ansermet, 
1211 Gen\`eve 4, Switzerland} 

\date{\today}

\def\openone{\leavevmode\hbox{\small1\kern-3.3pt\normalsize1}}

\newcommand{\la}{\langle}
\newcommand{\ra}{\rangle}
\newcommand{\be}{\begin{equation}}
\newcommand{\ee}{\end{equation}}
\newcommand{\bea}{\begin{eqnarray}}
\newcommand{\eea}{\end{eqnarray}}
\newcommand{\pd}{\partial}

\newcommand{\APPROX}[1]{            
   {{\raisebox{-.3cm}{$\textstyle\sim$}} \atop {\scriptstyle{#1}}}}
\begin{document}

\twocolumn[\hsize\textwidth\columnwidth\hsize\csname@twocolumnfalse\endcsname
\maketitle

\begin{abstract}
We investigate fractional Brownian motion with  a microscopic random--matrix model and  introduce a fractional Langevin equation. We use  the latter   to study  both sub- and superdiffusion of a  free particle coupled to  a fractal heat bath. We further compare fractional Brownian motion with  the fractal time process. The  respective mean--square displacements of  these two forms of anomalous diffusion exhibit  the same power--law behavior. Here we show that their lowest  moments are actually all identical, except  the second moment of the velocity. This provides a simple criterion which enables to  distinguish  these two non--Markovian processes.
\end{abstract}
\pacs{PACS numbers: 05.40.Jc, 05.30.-d}
\vskip.5pc]
Diffusion is one of the  basic non--equilibrium phenomena. Normal diffusion is well described in the theory of Brownian motion as a Gaussian  process that  is both local in space and in time. It is characterized by a mean--square displacement which is asymptotically linear in time, $\la x^2 \ra =2Dt$, where $D$ is the diffusion constant \cite{ris89}. However, a growing number of experimental observations show that more complex processes, in which the mean--square displacement is not proportional to $t$, also occur in nature.  Anomalous diffusion has for instance been seen in micelle systems \cite{ott90}, in two--dimensional  rotating flows \cite{sol93}, in porous glasses  \cite{sta95}, in actine networks \cite{amb96}, but also   on capillary surface waves \cite{han97}, in strongly coupled dusty plasmas \cite{jua98}, and more recently in  intracellular transport \cite{cas00}. Anomalous diffusion finds its dynamical origin in non--locality, either in space or in time. A well--known example of a process which is non--local in space  is L\'evy stable motion,  for which   the mean--square displacement  is actually infinite due to the occurrence of very long jumps \cite{lut01a}.  In this Letter we focus  on processes which are non--local in time  and whence show memory effects.  Specifically, we shall discuss and compare  fractional Brownian motion (fBm)  \cite{man68} and the fractal time  process (ftp)\cite{mon87}. These two forms of anomalous diffusion are fundamentally different (see  below).  Yet,  they are difficult to tell apart  experimentally, since  both yield  a mean--square displacement  of the form $\la x^2 \ra \propto t^\alpha$, $\alpha\neq1$.  It  is for instance still an open question whether the long--range correlations observed  in nucleotide sequences \cite{li92,vos92,pen92} are to be interpreted in terms  of fBm or ftp type DNA walks \cite{all98}.  In this paper we aim at providing a simple criterion which permits  to distinguish between  these two non--Markovian processes. 

 The very difference  between fBm and ftp is best illustrated by looking at their diffusion equations.  The solution of the  diffusion equation for fBm  \cite{wan90}
\be
\label{eq1}
\frac{\pd}{\pd t} P_{\mbox{\tiny fBm}}(x,t) =  \alpha D t^{\alpha-1} \frac{\pd^2}{\pd x^2}P_{\mbox{\tiny fBm}}(x,t)
\ee
is  easily found to be the Gaussian distribution 
$P_{\mbox{\tiny fBm}}(x,t)  = \exp (-x^2/4Dt^\alpha)/[4\pi Dt^\alpha]^{1/2}$.  FBm thus describes  Gaussian transport. It is important to note that  Eq.~(\ref{eq1}) is thereby local in time (there is no memory kernel). The non--Markovian character is expressed through a time--dependent diffusion constant,  $D_\alpha(t) =  \alpha D \, t^{\alpha-1}$. In contradistinction,  the  diffusion equation for  ftp\cite{bal85}
\be
\label{eq2}
\frac{\pd}{\pd t}P_{\mbox{\tiny ftp}}(x,t)=  \frac{D }{\Gamma(\alpha-1)}\int_0^t \frac{d\tau}{(t-\tau)^{2-\alpha}}\frac{\pd^2 }{\pd x^2}P_{\mbox{\tiny ftp}}(x,\tau) 
\ee
contains a memory kernel and the distribution function  $P_{\mbox{\tiny ftp}}(x,t)$ is hence non--Gaussian. The solution of (\ref{eq2})  is given  by  $ P_{\mbox{\tiny ftp}}[x,z]  = \exp (-|x|z^{\alpha/2}/D^{1/2})/[ 2 D^{1/2} z^{1-\alpha/2}]$ in Laplace $z$ space. In time,  $P_{\mbox{\tiny ftp}}(x,t)$ has  been expressed in closed form in terms of a Fox function \cite{sch89} or a one sided L\'evy stable distribution \cite{bar00a}. By introducing   further the Riemann--Liouville fractional derivative ($0\!<\!\lambda\!<\!1$) \cite{sai97}
\be
\label{eq3}
\frac{\pd^\lambda f(t)}{\pd t^\lambda} = \frac{1}{\Gamma(-\lambda)} \int_0^t \frac{f(\tau)\,d\tau}{(t-\tau)^{\lambda+1}} \ ,
\ee
 Equation (\ref{eq2})  can be rewritten as a fractional diffusion equation \cite{rem1}
 \be
\label{eq4}
\frac{\pd}{\pd t}P_{\mbox{\tiny ftp}}(x,t)= D  \frac{\pd^{1-\alpha}}{\pd t^{1-\alpha}}\frac{\pd^2 }{\pd x^2}P_{\mbox{\tiny ftp}}(x,\tau) \ .
\ee 
Both Eqs.~(\ref{eq1}) and (\ref{eq4}) reduce to the normal diffusion equation when $\alpha\!=\!1$.

We begin our discussion of fBm  by introducing a fractional Langevin equation. It is worthwhile to point out  that  the Langevin and the phase--space descriptions of Brownian motion are no longer fully equivalent  in the non--Markovian regime of interest here. As recently discussed by Calzetta et al.   \cite{cal00}, the Langevin equation contains more information and thus appears  more fundamental.  We then apply this fractional Langevin equation to study in some detail the anomalous diffusion of  a free particle coupled to a fractal heat bath. In particular, we  evaluate  the first two moments of both the  position and the velocity of the particle, which we express in terms of Mittag--Leffler functions.   Finally, we compare with  the results  obtained recently for ftp    by Metzler and Klafter for $0\!<\!\alpha\!<\!1$ \cite{met00} and by Barkai and Silbey for $1\!<\!\alpha\!<2$ \cite{bar00} by using a fractional Klein--Kramers equation. We find  that fBm and ftp  satisfy  the same generalized Einstein relation. Moreover,  their lowest   moments are all equal, except  the second moments of the velocity.

We  examine  the  dynamics of the Brownian  particle with  a microscopic  random--matrix model.  Random--matrix theory   has already been successfully applied in the context of 
 anomalous diffusion in  Refs.~\cite{kus97,kus99,lut01}. We thus consider   a system $S$   weakly coupled  to a fractal heat bath $B$ via a random--matrix interaction \cite{lut01,lut99}. The coupling is chosen linear in the position $x$ of  the system. The generic form of the Hamiltonian is  given by
\be
 \label{eq5}
H = H_S\otimes \openone_B + \openone_S\otimes H_B + x\otimes V \ ,
\ee
where $H_S= p^2/2M+U(x)$ is the  Hamiltonian of the system, $H_B$ describes the bath and $V$ is a centered Gaussian random band--matrix. It is assumed  that initially the system and the bath are uncorrelated and that the latter is in thermal equilibrium at temperature $\beta = (kT)^{-1}$. The variance of the random interaction is further taken   to have  the form \cite{lut01}
\be
\label{eq6}
\overline{{V_{ab}}^2} = A_0 \frac{|\varepsilon_a - \varepsilon_b|^{\alpha-1}}{\left[ \rho(\varepsilon_a)
    \rho(\varepsilon_b)\right]^{\frac{1}{2}}}\, 
    \exp\left[-\frac{(\varepsilon_a - \varepsilon_b)^2}{2\Delta^2}\right] \ .
\ee
Here $\varepsilon_a$'s denote the eigenenergies of the bath Hamiltonian
($H_B|a\ra = \varepsilon_a |a\ra$), $A_0$ is the strength of the
coupling, $\Delta$  the bandwidth and $\rho(\varepsilon)$ is the density
of states of the bath, which is locally written as $\rho(\varepsilon)= \rho_0 \exp(\beta \varepsilon)$. As shown in \cite {lut01}, the variance   (\ref{eq6}) gives rise to subdiffusion when $\alpha\!<\!1$ and to superdiffusion when $1\!<\!\alpha\!<\!2$.  The coupling to the bath is characterized by the bath correlation function which is defined as $ K(t)= \la \widetilde{V}(t)\widetilde{V}(0) \ra_{B}= K'(t) + i K''(t)$. Here $\widetilde{V}(t)=\exp(iH_B t)\, V \exp(-iH_B t)$ 
and $\la\dots\ra_B  $
denotes the thermal average. After  performing the average over the random--matrix ensemble, $ \overline K(t)$  is found to be  simply the Fourier transform of the variance $\overline{{V_{ab}}^2}$ with respect to $\varepsilon_b$. In the following  we consider the limit of  high temperature and large bandwidth, $1\ll\Delta\ll kT$. Using  the variance  (\ref{eq6}) we then obtain
\be
\label{eq7}
\overline{K'}(t) =  2 A_0 \Gamma(\alpha) \cos(\frac{\alpha\pi}{2}) \,t^{-\alpha}\! \ ,\;\;\;\overline{K''}(t) =\frac{\beta}{2}  \frac{d \overline{K'}}{dt} \ .
\ee
We see that the time dependence of $\overline{K}(t)$  follows  an inverse power law. This presence of a long tail leads to long--time  correlation effects in the dynamics of the Brownian system \cite{lut01}.  Note that  for $\alpha\!=\!1$, the Fourier transform of  (\ref{eq6}) reads $\overline{K'}(t) =  2 \pi A_0 \,\delta(t)$ and normal Brownian motion is recovered.
Moreover,  in the limit of weak coupling considered  here, we have shown in \cite{lut01} that   the random band--matrix model can be mapped onto  the oscillator bath model \cite{gra88}: There is a one--to--one correspondence between the variance  $\overline{{V_{ab}}^2}$ of the random band--matrix model and the spectral density function $J(\omega)$ of the oscillator bath model.  The generalized Langevin equation corresponding  to the random--matrix Hamiltonian (\ref{eq5})  can therefore be readily written as \cite{gra88}
\be
\label{eq8}
M \ddot x(t) +M \int_0^t \gamma(t-\tau) \dot x(\tau) d\tau +U'(x)= \xi(t) \ ,
\ee
where $\xi(t)$ is a Gaussian random force with mean zero and variance $\la \xi(t)\xi(0)\ra=  \overline{K'}(t)$, and $\gamma(t)$ is a damping kernel  which obeys $M kT \gamma(t) = \overline{K'}(t)  $. This last relation is often referred to as the second fluctuation--dissipation theorem \cite{kub66}. Remark that the Langevin equation is completely determined by the real part $\overline{K'}(t)$ of the bath correlation function. Furthermore, in the limit of weak coupling, the dynamics described by the Hamiltonian  (\ref{eq5})  is Gaussian and one can show that the corresponding diffusion equation is precisely given by Eq.~(\ref{eq1}). Using again  the fractional derivative (\ref{eq3}), we may rewrite 
Eq.~(\ref{eq8}) in the form of a fractional Langevin equation. We obtain
\be
\label{eq9}
M \ddot x + M  \gamma_\alpha  \,\frac{\pd^{\alpha-1}}{\pd t^{\alpha-1}}\dot x(t) +U'(x) = \xi(t) \ ,
\ee
where we have defined  $\gamma_\alpha =  \pi A_0 \beta  /(M  \sin(\alpha\pi/2) )$.  The fractional Langevin equation  (\ref{eq9}) describes both subdiffusion for $0<\!\alpha\!<1$ and  superdiffusion for  $1<\!\alpha\!<2$ \cite{rem2}. 
As a simple application of the fractional  equation (\ref{eq9}), we now concentrate on  the free particle and accordingly set  $U(x) =0$. In this case, the solution of the Langevin equation  is   easily obtained by applying  Laplace transform techniques \cite{por96}.  We find 
\be 
\label{eq10}
x(t) = x_0 + v_0   B_v(t) + \int_0^t  B_v(t-\tau) \xi(\tau) d\tau\ ,
\ee
where $(x_0,v_0)$ are the initial coordinates of the particle and $B_v(t) = \int_0^t C_v(t') dt'$ is the integral  of the (normalized) velocity autocorrelation function  $C_v(t) = \la v(t) v\ra/ \la v^2\ra$. The  Laplace transform of $C_v(t)$ is given by 
\be 
\label{eq11}
C_v[z] =  \frac{1}{z+\gamma[z]} = \frac{1}{z+\gamma_\alpha z^{\alpha-1}} \ ,
\ee
where $\gamma[z]$ is the Laplace transform of the damping kernel.
Eq.~(\ref{eq11})  is  known as the first fluctuation--dissipation theorem \cite{kub66}. By taking the  inverse Laplace transform, the velocity autocorrelation function  can be written as
\be
\label{eq12}
 C_v(t) = E_{2-\alpha}(-\gamma_\alpha t^{2-\alpha}) \ .
\ee
Here we have introduced the Mittag--Leffler function $E_\alpha(t) $, which is  defined by the series expansion \cite{erd54}
\be
\label{eq13}
E_\alpha(t) = \sum_{n=0}^\infty \frac{t^n}{\Gamma(\alpha n+1)} \ .
\ee 
The function  $E_\alpha(t) $ reduces to the exponential when $\alpha=1$.
The asymptotic behavior of  the Mittag--Leffler function (\ref{eq13})  for short and long times is  respectively given by $\sim \exp(t)$ and $\sim -(t\,\Gamma(1-\alpha))^{-1}$. For the velocity autocorrelation function (\ref{eq12}) this yields a typical stretched exponential behavior at short times
\be
 \label{eq14}
C_v(t) \sim \exp\frac{-\gamma_\alpha t^{2-\alpha}}{\Gamma(3-\alpha)}, \hspace{1cm} t \ll 1/(\gamma_\alpha)^{\frac{1}{\alpha}} \ ,
\ee
and an inverse power--law tail at long  times
\be
\label{eq15}
C_v(t) \sim \frac{t^{\alpha-2}}{\gamma_\alpha \Gamma(\alpha-1)} , \hspace{1cm} t \gg\frac{1}{(\gamma_\alpha)^{1/\alpha}} \  .
\ee
The result (\ref{eq15}) has already been derived  in Ref.~\cite{lut01}, where it has  been shown to induce the ``whip--back'' effect.  After time integration, we finally get from Eq.~(\ref{eq12})
\be 
 B_v(t) = t \,E_{2-\alpha,2}(-\gamma_\alpha t^{2-\alpha}) \ ,
\ee
where  we have used the generalized Mittag--Leffler function  $E_{\alpha,\beta}(t)$  defined as \cite{erd55}
\be 
E_{\alpha,\beta}(t) = \sum_n ^\infty \frac{t^n}{  \Gamma( \alpha n+\beta)}\  .
\ee
In the  long--time limit, the generalized Mittag--Leffler function satisfies $E_{\alpha,\beta}(t) \sim -(t\, \Gamma(\beta-\alpha))^{-1}$.  Accordingly,  $B_v(t)$ exhibits a decay of the form 
\be
B_v(t)\sim \frac{t^{\alpha-1}}{\gamma_\alpha \Gamma(\alpha)}\ , \hspace{0.7cm}\mbox{ when }  t\rightarrow \infty \ .
\ee
We emphasize  that  the solution (\ref{eq10}) of the fractional Langevin equation in   the force  free case is completely specified by  the knowledge of the function  $B_v(t)$  .

Let us  now turn to the evaluation of the lowest moments of the position and the velocity of the free particle.
The mean displacement and the mean--square  displacement  are readily deduced from Eq.~(\ref{eq10}). We find 
\be
\label{eq16}
\la x \ra = x_0 + v_0 \,  t \,E_{2-\alpha,2}(-\gamma_\alpha t^{2-\alpha}) \APPROX{t\rightarrow \infty}  \frac{v_0}{\gamma_\alpha} \frac{t^{\alpha-1}}{\Gamma(\alpha)}
\ee
and 
\bea
\label{eq17}
\la x^2 \ra \,& & = \frac{2kT}{M} t^2 E_{2-\alpha,3}(-\gamma_\alpha  t^{2-\alpha})\nonumber\\
& & \APPROX{t\rightarrow \infty}\frac{2kT}{\gamma_\alpha M} \frac{t^\alpha}{\Gamma(1+\alpha)}\ .
\eea
In the last equation, thermal initial conditions have been assumed ($x_0 =0$, $v_0^2 = kT/M$). In addition, one may easily verify  that $\la x \ra^2$, $\la x^2 \ra $ and $C_v(t)$ satisfy the general Green--Kubo relation
\be
\label{eq18}
\la x^2 \ra -\la x\ra^2= \frac{2kT}{M} \int_0^t dt'\int_0^{t'} d\tau \, C_v(\tau) \ ,
\ee
which is known from linear response theory \cite{kub66}.
In a similar way, one can compute the first and second moments of the velocity from the time derivative of Eq.~(\ref{eq10}). This results in
\be
\label{eq19}
\la v\ra = v_0\, E_{2-\alpha}( -\gamma_\alpha t^{2-\alpha})   \APPROX{t\rightarrow \infty}\frac{v_0}{\gamma_\alpha} \,t^{\alpha-2}\ ,
\ee
and
\bea
 \label{eq20}
\la v^2\ra =&& v_0^2 \Big[ E_{2-\alpha}(-\gamma_\alpha t^{2-\alpha})\Big]^2 \nonumber \\
                  && + \frac{kT}{M} \Big\{ 1- \Big[ E_{2-\alpha}(-\gamma_\alpha t^{2-\alpha}) \Big]^2 \Big\} \ .
\eea
We observe that $\la v^2\ra$   decays like $(t^{\alpha-2})^2$ for large $t$.  A common remarkable property of the above calculated mean values is their slow relaxation towards equilibrium as given by the (generalized) Mittag--Leffler function. This has to be contrasted with normal Brownian motion where all this quantities display  an  exponential decay. Let us now  discuss  the generalized Einstein relation  which relates driven and free process \cite{met99}. We consider a particle initially at rest  ($x_0\!=\!v_0\!=0$) and seek the mean position $\la x \ra_F$ as a function of an externally applied constant force $U(x)= -x F\theta(t)$.  From  the Langevin equation we easily find 
\be
\label{eq21}
\frac{d \la x \ra_F}{dt} = \frac{F}{M} \int_0^t C_v(t') dt'\  ,
\ee
where the velocity autocorrelation function $C_v(t)$ is given by Eq.~(\ref{eq12}).
Equation (\ref{eq21})  together with the  Green-Kubo relation (\ref{eq18}) for $\la x^2\ra_0$ in the force free case, then yields the generalized Einstein relation for fBm
\be
\label{eq22}
\la x \ra_F = \frac{F}{2kT} \, \la x^2\ra_0 \ .
\ee
It is interesting to note that the validity of the  Einstein relation (\ref{eq22}) has been recently verified experimentally \cite{bar98,gu96}. 

We now come  to the comparison  of  fBm with ftp.  Barkai and Silbey have investigated superdiffusive ftp with a  fractional Klein--Kramers equation they inferred from  a generalized Rayleigh model \cite{bar00}.  For the free particle, a  direct comparison \cite{rem3} between their results  and our Eqs.~(\ref{eq16})--(\ref{eq22}) shows that  the mean displacement (\ref{eq16}), the mean--square displacement (\ref{eq17}),  the  velocity's first moment (\ref{eq19}) and the velocity autocorrelation function (\ref{eq12}) are identical for the two processes. This means in particular that fBm and ftp satisfy the same Green--Kubo  relation (\ref{eq18}). Moreover, both fBm and ftp obey  the same  generalized Einstein relation (\ref{eq22}). Although fBm and ftp are fundamentally different processes, we thus notice that  they  share strikingly common features. However, the second moments  of the velocity are different. For convenience, we  quote   their equation (2.18) which reads (in our notation)
\bea
 \label{eq23}
\la v^2\ra _{\mbox{\tiny ftp}}=&& v_0^2  \, E_{2-\alpha}(-2 \gamma_\alpha t^{2-\alpha})\nonumber \\
                  && + \frac{kT}{M} \Big\{ 1-  E_{2-\alpha}(-2\gamma_\alpha t^{2-\alpha})  \Big\} \ .
\eea
 We  see that for ftp, the second moment of the velocity relaxes  asymptotically  like $t^{\alpha-2}$. This is in sharp contrast to the fBm result Eq.~(\ref{eq20}) which exhibits a much faster decay.  It is also worth  noting  that Eqs.~(\ref{eq19}) and (\ref{eq23}) reduce to the same (exponential) expression for $\alpha=1$. On the other hand, subdiffusive ftp has been studied by Metzler and Klafter by using a fractional Klein--Kramers equation derived from a non--Markovian  generalization of the  Chapman--Kolmogorov equation.  A comparison with  their results for the force free case leads to similar conclusions as in the superdiffusive regime. Many experiments on  anomalous diffusion have measured either the mean--square displacement \cite{amb96,han97,jua98,cas00} or the  generalized Einstein relation \cite{bar98,gu96}.  However, the latter do not allow to distinguish  fBm and ftp, as we have just shown. In contrast,  the variance of the velocity offers a clear distinction between these two processes as exemplified by Eqs.~(\ref{eq20}) and (\ref{eq23}). This is the main result of this Letter. 

In summary, we have investigated fBm within a random--matrix approach and introduced a fractional Langevin equation which applies for both sub- and superdiffusion. We have studied the anomalous dynamics of a free particle coupled to a fractal heat bath and performed a comparison between fBm and ftp.  We have found that these completely different forms of non--Markovian anomalous diffusion share many common characteristics. In particular, they satisfy the same generalized Einstein relation and their lowest moments are all equal with the exception of the second moment of the velocity.

We thank E. Barkai for reading the manuscript.


\begin{references}
\bibitem{rem1} Here, for simplicity, we only consider  $\alpha<1$.
\bibitem{rem2}  To avoid difficulties when  $1\!<\!\alpha\!<\!2$, the friction term should be integrated by part to become  \be
M \gamma_\alpha (1-\alpha)\Big[\frac{v_0 t^{\alpha-1}}{\Gamma(1-\alpha)} + (1-\alpha)  \frac{\pd^{\alpha-2}}{\pd t^{\alpha-2}} \ddot x(t)\Big] \ .
\ee
\end{references}
\end{document}